\documentclass[twocolumn, a4paper, 10pt]{IEEEtran}
\usepackage{graphics}
\usepackage{amsmath}
\usepackage{url}
\makeatletter

\usepackage{epsfig}
\usepackage{subfigure}
\usepackage{graphicx}
\usepackage{algorithm}
\usepackage{algorithmic}
\usepackage{times}
\usepackage{verbatim}
\def\unnumfootnote{\xdef\@thefnmark{}\@footnotetext}

\makeatother
\begin{document}
\title{Topological Analysis of the Power Grid and Mitigation Strategies Against Cascading Failures}
\author{\itshape{S.Pahwa}\footnotemark${^{1}}$, A.Hodges\footnotemark${^{1}}$, C.Scoglio\footnotemark${^{1}}$, S.Wood\footnotemark${^{2}}$}
\maketitle
\thispagestyle{empty} 
\footnotetext[1]{Kansas State University, Manhattan, KS, USA}
\footnotetext[2]{Georgia Institute of Technology, Atlanta, Georgia, USA}

\unnumfootnote{Received : November 1, 2009; Revised Version : March 5, 2010.\\ This work is supported by the Energy and Power Affiliates Program, consisting of Westar Energy, Burns and McDonnell, Omaha Public Power District, and Nebraska Public Power District.}
\begin{abstract}
This paper presents a complex systems overview of a power grid network. In recent years, concerns about the robustness of the power grid have grown because of several cascading outages in different parts of the world. In this paper, cascading effect has been simulated on three different networks, the IEEE 300 bus test system, the IEEE 118 bus test system, and the WSCC 179 bus  equivalent model. \emph{Power Degradation} has been discussed as a measure to estimate the damage to the network, in terms of load loss and node loss. A network generator has been developed to generate graphs with characteristics similar to the IEEE standard networks and the generated graphs are then compared with the standard networks to show the effect of topology in determining the robustness of a power grid. Three mitigation strategies, Homogeneous Load Reduction, Targeted Range-Based Load Reduction, and Use of Distributed Renewable Sources in combination with Islanding, have been suggested. The Homogeneous Load Reduction is the simplest to implement but the Targeted Range-Based Load Reduction is the most effective strategy.\\ \\
\end{abstract}
\section{Introduction}
It is well-known that power grids are among the largest and most complex technological systems ever developed~\cite{MKA:07}. The power grid can be represented by a large graph belonging to a special family of graphs called complex networks. As is the case with most networks, the edges are not costless~\cite{LAMH:00}, so there is no node or group of nodes that prevail over the others, and the distribution can be fitted by an exponential function~\cite{PVM:04}. However, if we calculate the load on each node, we observe that although the network is not very heterogeneous in the node degree, it shows high heterogeneity in the node load. Most of the nodes handle a small load but there are a few nodes that have to carry an extremely high load~\cite{PVM:04}. The same is true for links also. Thus, some nodes and links tend to become more important than the others and an intentional or accidental removal of these elements can lead to the collapse of the entire network. Such incidents have taken place in history, such as the one on August 10, 1996 when a 1300 MW electrical line in Southern Oregon sagged in summer heat, initiating a chain reaction that cut power to more than four million people in eleven Western States~\cite{BDIA:00}, ~\cite{MBV:00}. Another example is the incident of August 14, 2003 when an initial disturbance in Ohio~\cite{JR:03} led to the largest blackout in the history of the United States and millions of people throughout parts of  North Eastern and Mid Western United States, and Ontario, Canada, were without power for as long as fifteen hours.\\
North American Electric Reliability Corporation (NERC) defines a cascading failure as ``The uncontrolled loss of any system facilities or load, whether because of thermal overload, voltage collapse, or loss of synchronism, except those occurring as a result of fault isolation~\cite{NERC:05}."
The power grid network is different from other complex networks in that the flow dynamics depend greatly on the electrical characteristics, such as the impedances of the transmission lines. Power flows along the least resistive path, and thus the amount of power flowing through the transmission lines is inversely proportional to their impedances.\\
If a single line gets overloaded or breaks, its power is immediately shifted to a different line and the disturbance can be suspended. However, if the redistribution of power leads to the subsequent overloading of other lines, the consequence could be a cascade of overloading failures.\\
In section II, we discuss research and work related to the topological analysis of the power grid. In section III, we describe the simulator that we developed to study the cascading effect on power grid networks. We analyze the IEEE 300 bus test system~\cite{WAS}, the IEEE 118 bus test system~\cite{WAS}, and the WSCC 179 bus equivalent system for cascading failures using the DC Power Flow Model~\cite{BR:88}. The redistribution of power is dependent on the electrical characteristics, such as impedances, of transmission lines. The resistances of the lines have been neglected because they are very small as compared to their inductive reactances~\cite{BR:88}. In section IV, we discuss a Network Generator that we developed to generate a family of power grid networks with characteristics such as number of nodes, average node degree, and maximum node degree, same as the standard networks but with random connections. We then compare the different network metrics of the standard and generated graphs to determine the effect of topology on their robustness. We investigate mitigation strategies for  cascading failures in Section V. The results, and conclusions, along with future work have been presented in sections VI and VII, respectively. 
\section{Related Work}
The United States Department of Energy states that the future of the power grid in the United States includes expansion, updating, and implementation of the Smart Grid~\cite{KD:09}, along with the improvement in robustness of the already existing network. The complete implementation of the smart grid would bring about a revolution in the way the power industry works today. Distributed renewable energy will be an integral part of the smart grid~\cite{KD:09} to help meet the demand of future energy needs. Current research is geared towards the study of renewable energy and its use to improve the efficiency of the smart grid. While the use of technology and advanced communication methods is one of the ways to improve the resilience and robustness of the power grids, modeling the grid as a complex system and analyzing its topology to determine robustness is also a way. Methods for modeling the power grid as a complex network have developed to a point where the vastness and complexity of the grid can be realized in a simulation~\cite{ASB:08, MMFAS:07}.\\
The authors of~\cite{AY:00} have mentioned that energy travels between a set of nodes using the shortest path and that the load on a node is the total number of shortest paths passing through that node. As a result, the load distribution changes with change in the network, with ER random graphs having an exponential load distribution and BA networks exhibiting a power law distribution in the node load. The analysis is completely based on node capacity without any mention of the link (transmission line)characteristics and is based on single node failures.\\
In~\cite{JJI:05}, the authors discuss a hidden-failure embedded transmission network model to explore the characteristics of cascading events. The impacts of various model parameters on the system dynamics are examined, and the corresponding mitigation strategies are also suggested. They suggest that the cascading failures arising due to single element failure are actually supplemented by hidden failures, which are exposed only in case of faults.\\
In this paper, we analyze power grid networks with respect to redistribution of flows, and failure of links due to overloading. We compare the characteristics of IEEE standard networks with generated networks to show how topology can affect the robustness of a network. The generated graphs are obtained by keeping the number of nodes, average node degree, and maximum node degree same as the standard graphs but by connecting links randomly. Keeping the topology same as the standard, some mitigation strategies can be applied to the networks to reduce cascading effects. We have discussed such mitigation strategies, based on load reduction, for the power grid networks.

\section{Cascading Effect in a Power Grid}
To analyze the effect of topological characteristics of the power grid network on cascading failures, we have used networks such as the IEEE 300 bus test system~\cite{WAS}, the IEEE 118 bus test system~\cite{WAS}, and the WSCC 179 bus equivalent system. The buses in the power grid are referred to as nodes, the transmission lines as links, and the impedances of the transmission lines as weights on the links. The 300 node network consists of 247 nodes plus two smaller subsections of nodes which are not well connected with the main graph. These two subsections were not critical for the analysis from a topological view point and thus were not included. Therefore, the 300 node network will be referred to as the 247 node network here onwards unless we refer to the standard test case. After the initial failure, some of the links get overloaded and fail. This represents the first stage of cascade. The first stage leads to further overloading of more links and their collapse, constituting the second stage and so on. In this way, the systems go through multiple stages of cascade before they finally stabilize and there are no more failures. We categorized the links into two types - vulnerable and non-vulnerable - depending upon whether they cause more or less than ten percent damage upon removal. Damage represents both, the loss in connectivity and the loss of load. Approximately $41.97\%$ of the links in the 247 nodes network were found to be vulnerable and the remaining $58.03\%$ were non-vulnerable links. We use Power Degradation as a measure to determine the severity of damage to the network as a result of initial failure. Power Degradation is the fall in the total load of the system as compared to the original load. We also compare the networks in terms of the number of nodes lost by the failure of a vulnerable link. 
\begin{figure}[h]
	\includegraphics[width = 3in,  height = 2in]{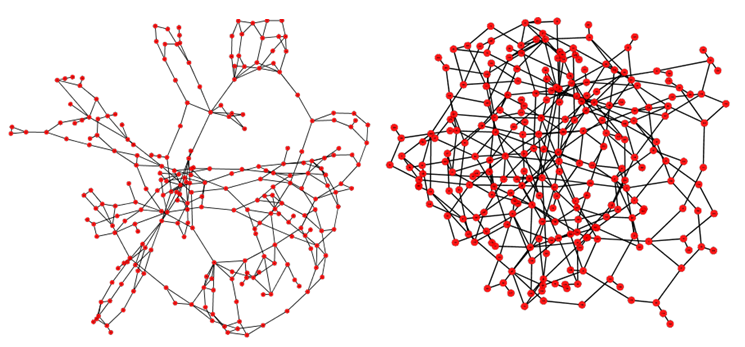}
	\caption[center]{247 Nodes Original (left) and Generated Networks}
	\label{Fig1}
\end{figure} 
\begin{figure}
	\includegraphics[width = 3in,  height = 2in]{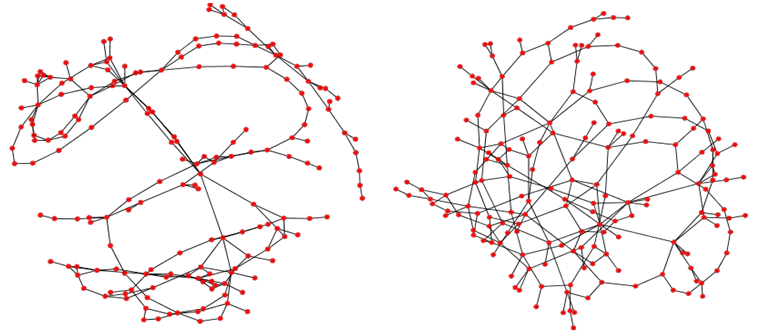}
	\caption[center]{179 Nodes Original (left) and Generated Networks}
	\label{Fig2}
\end{figure}
\section{Network Generation and Analysis}
\begin{figure}
	\includegraphics[width = 3in, height = 2in]{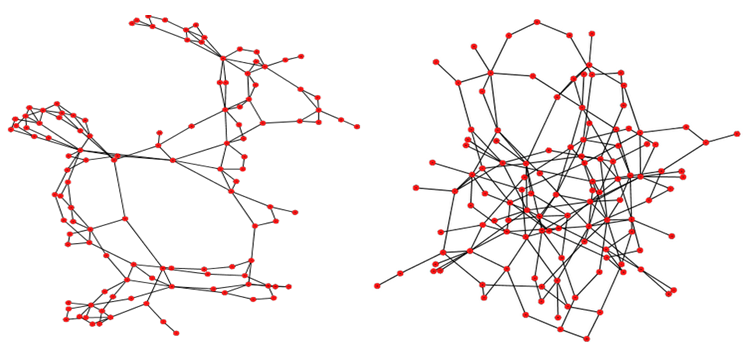}
	\caption[center]{118 Nodes Original (left) and Generated Networks}
	\label{Fig3}
\end{figure}

\begin{figure}
		\includegraphics[width = 3.5in, height =  1.5in]{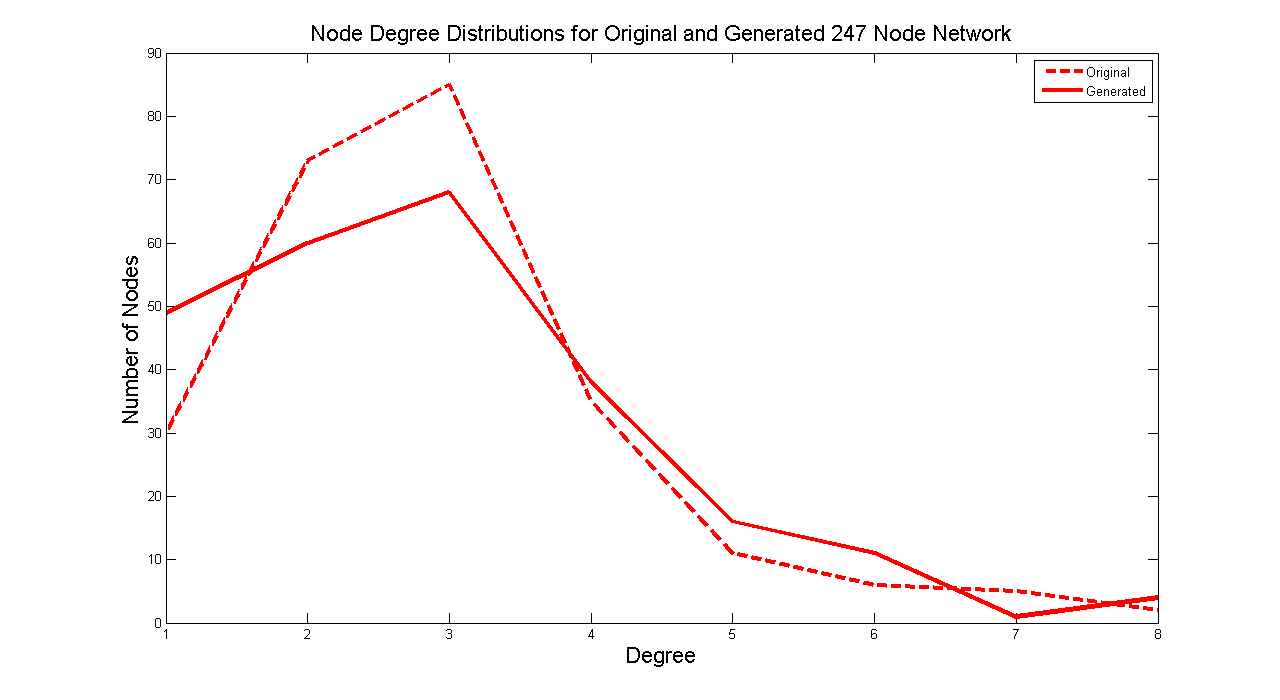}
	\caption[center]{Node Degree Distribution of the 247 Nodes Original and Generated Networks}
	\label{Fig4}
\end{figure}

\begin{figure}
		\includegraphics[height = 2in]{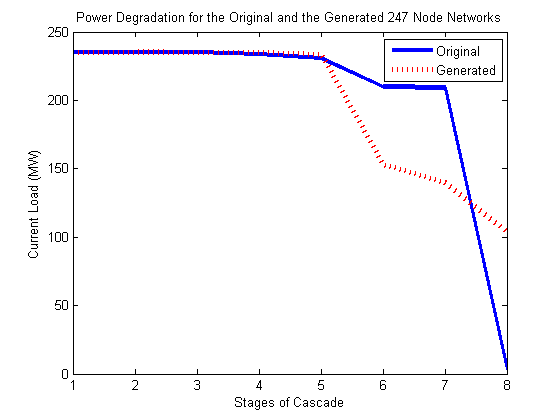}
	\caption[center]{Power Degradation Graphs for the 247 Nodes Original and Generated Networks}
	\label{Fig5}
\end{figure}

\begin{figure}
		\includegraphics[height = 2in]{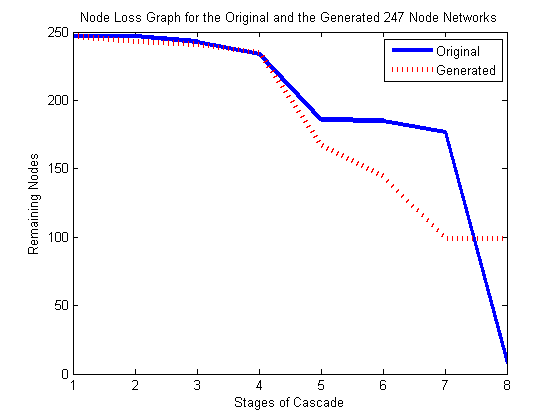}
	\caption[center]{Node Loss Graphs for the 247 Nodes Original and Generated Networks}
	\label{Fig6}
\end{figure}

To analyze the effect of topological characteristics of power grid networks on cascading effect, we use the standard networks such as the IEEE 300 bus~\cite{WAS}, IEEE 118 bus ~\cite{WAS}, and WSCC 179 bus networks as the first stage and study cascading failures on them. Then we design a first approximation network generator to produce networks having characteristics like number of nodes, maximum node degree, and average node degree similar to the original networks. Fig. \ref{Fig1}, \ref{Fig2}, and \ref{Fig3} show the original and generated networks for the 247 node, the 179 node, and the 118 node systems, respectively. Fig. \ref{Fig4} shows the node degree distributions of the 247 node original and generated networks. This figure shows that the generated network follows the node degree distribution of the original network closely. The horizontal axis represents the degree from 1 to the maximum node degree, which is 8 in case of the 247 node network, and the vertical axis represents the number of nodes with a particular node degree. The generated graph is one of those from the family of random graphs having the node degree distribution similar to the original network. The generator takes the number of nodes, the maximum node degree, and the average node degree as inputs and gives the edge list of the network as the output. The edge list is a list which contains pairs of vertices that every edge connects. These generated networks are then used for the simulation of cascading effect. We generated the weights and loads probabilistically from measured inductance and load distributions. Fig. \ref{Fig5} shows the graphs for power degradation on the 247 node original and generated networks. The horizontal axis represents the stages of cascade that the system goes through, before it finally stabilizes, and the vertical axis is the current load on the system in megawatt. An important observation to be made here is that as the load-carrying nodes get disconnected from the network, the total load on the network keeps reducing. As indicated by Fig. \ref{Fig5}, the original network loses nodes gradually, and so stabilization is achieved at a much later stage than the generated network, which loses the excess load quickly. Thus, this improvement in the amount of power degradation can be safely attributed to the generated topology of the power grid. Fig. \ref{Fig6} shows the graph for node loss in the 247 node original and generated networks. The horizontal axis again represents the stages of cascade and the vertical axis represents the number of nodes remaining in the network, as a result of cascading.
Table \ref{tab:Table1} shows the comparison between different characteristics of the original and the generated networks.
\begin{table}
\caption[center]{Differences in Characteristics of Original and Generated Networks}
		\begin{tabular}[width=0.5in]{|c|c|c|c|c|}
		\hline Network&Nodes&Char Path Length&Clustering Coefficient\\
		\hline 
		\hline &$247$ & $9.646$ & $0.102$\\
					 Original &$179$ & $12.382$ & $0.089$\\
					 &$118$ & $6.309$ & $0.165$\\
		\hline &$247$ & $5.361$ & $0.004$\\
					 Generated &$179$ & $5.549$ & $0.010$\\
					 &$118$ & $4.385$ & $0.020$\\
		\hline
		\end{tabular}
	\label{tab:Table1}
\end{table}
The table indicates that the generated networks have a shorter characteristic path length as compared to the original. This metric causes the generated networks to be better connected and hence contribute to the robustness of the networks. However, the clustering coefficient of the generated networks is lower than that of the original networks due to the more random nature of the graph in the generated case. Therefore, we can say that there is less local connectivity due to smaller clustering coefficient but higher degree of global connectivity due to shorter characteristic path length.
\section{Mitigation Strategies}
We propose the following mitigation strategies to limit the damage to the network by cascading failures: Homogeneous Load Reduction, Targeted Range-Based Load Reduction, and Islanding with use of Distributed Renewable Sources. The third strategy aims at optimally selecting clusters or islands of nodes that can be separated from the main grid and be independently powered using distributed sources such as wind turbines. Each of these strategies is discussed in detail.
\subsection{Homogeneous Load Reduction}
\begin{figure}
	\includegraphics[height = 2in]{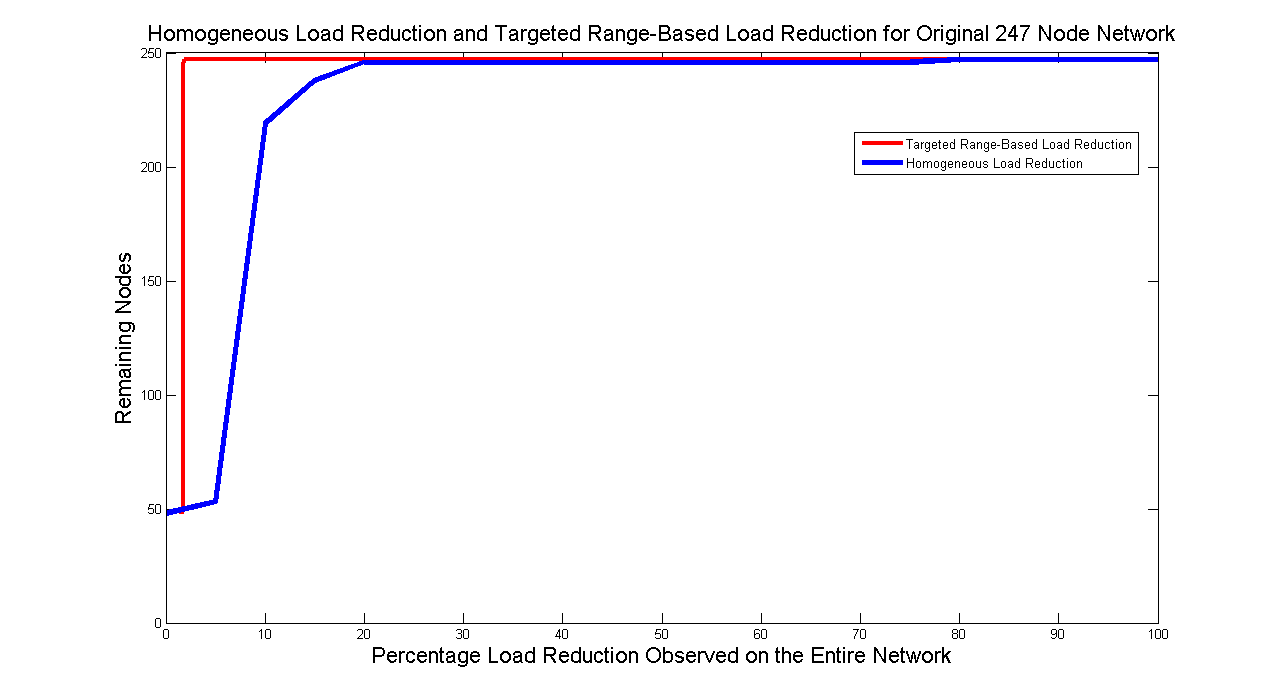}
	\caption[center]{Homogeneous Load Reduction and Targeted Range-Based Load Reduction}
	\label{Fig7}
\end{figure}

This mitigation strategy aims at reducing a given percentage of the load on each of the nodes in the network, after the initial failure. This reduction in load attempts to keep the nodes and links operating below their maximum capacities and to better accommodate the redistribution of load due to failure of links or nodes. We performed a series of simulations on the original 247 node network, in which the load on each of the nodes was reduced from hundred percent to zero percent, in steps of five. We plotted the result of each simulation for the original 247 node network to obtain the Homogeneous Load Reduction curve as shown in the Fig. \ref{Fig7}. Each point on the horizontal axis is a separate simulation with the starting load for each being the initial load on the nodes. As Fig. \ref{Fig7} shows, we get a connected component with most of the nodes within the component at about 10\% load reduction on all the nodes of the network. However, the total protection of the network is achieved at about 80\% reduction.
\subsection{Targeted Range-Based Load Reduction}
\begin{figure}[h]
	\includegraphics[height = 2in]{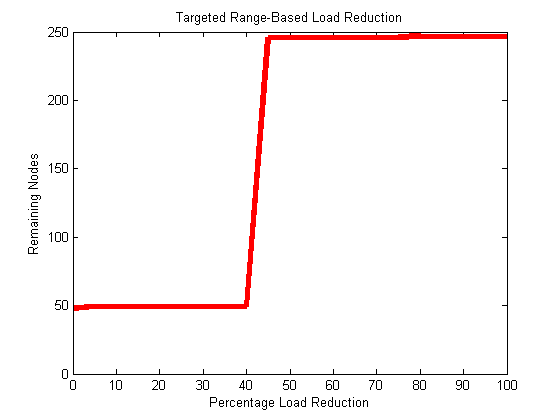}
	\caption[center]{Load Reduction on the Tree - Targeted Range-Based Load Reduction Strategy}
	\label{Fig8}
\end{figure}

If we select a node from the network and follow it along the adjacent nodes considering any one direction of power flow, we discover a tree for that node. The approach used in this strategy is to select one of the nodes connected to the link that failed, and discover the tree for that node. The node being considered is the one which was sending power out to the other node through the failed link. This particular node has to re-route the outgoing power among  its other connected neighbors. One or more of the neighbors of this root node may have their best way to the generator through it. These neighbors form a part of the tree. The same criterion is followed at every step until the entire tree is discovered. The tree stops either at a leaf node or at a source. Leaf nodes are nodes at the edge of the network. We then consider nodes in the tree which are not sources or which handle load, and reduce their load by a certain percentage. Since we consider a smaller set of nodes in this strategy, as compared to the Homogeneous Load Reduction, it constitutes a very small portion of the total load of the system, less than 2\%. Fig. \ref{Fig7} shows the comparison between the Homogeneous Load Reduction Strategy and the Targeted Range-Based Load Reduction strategy. 
Fig. \ref{Fig8} shows the load reduction only on the targeted tree. 
\subsection{Islanding with use of Distributed Renewable Sources}
The aim of this strategy is to disconnect a cluster or ``island" of nodes from the main grid to help reduce the load on the grid. This island will be powered by some distributed renewable sources such as wind turbines. The power output of each of the turbines is the theoretical load reduction on the transmission lines of the power grid. We are investigating some optimization techniques which could be used to disconnect the island from the main grid. 

\section{Results}
The following results were obtained from the discussions above:
\begin{itemize}
	\item The power degradation graphs in Fig. \ref{Fig5} show that the worst-case cascading effect stops at an earlier stage and causes less damage in the case of generated networks. The load loss in the case of the 247 node generated network discussed above is about 42.66\% less and the node loss is about 36.84\% less than the original network.
	\item As seen from Table \ref{tab:Table1}, the generated graphs have a smaller characteristic path length as compared to the real networks. This property of the generated networks makes them more robust against cascading failures. However, the clustering coefficient of the generated networks is lower than that of the original networks, due to more random nature of the generated networks.
	\item The Homogeneous Strategy is easier to implement, in that we simply reduce the load on all the nodes existing in the network. The Range-Based Load Reduction strategy is more economical, but requires cyber-physical system interaction to perform quick computations and reduce the load in only the desired regions. 
\end{itemize}
\section{Conclusions and Future Work}
The topology of the power grid network greatly contributes to its robustness. The topology determines the connectivity of the network and hence the number of alternate paths that can be taken by the network flow. The generated graphs are better connected because of shorter characteristic path lengths and are more random in nature than the real power grid networks. As a result of randomness, they show more robustness against cascading failures. However, the feasibility of having shorter characteristic path lengths in a real power grid network, must be investigated. The distributed renewable sources can be of great assistance in reducing the ever-increasing load on the power grid.
Our future work includes the following:
\begin{itemize}
	\item Using optimization techniques to obtain feasible points for disconnection of node clusters for Islanding, and
	\item Weighted Analysis of the power grid network to study the deviation of the optimal paths from the shortest paths. 
\end{itemize}
\section{Acknowledgment}
The authors would like to thank A. Pahwa, S. Starrett, I. Dobson, and P. Schumm for their valuable suggestions and comments.
\bibliographystyle{IEEE}
\bibliography{PG-bib}
\end{document}